\documentclass[sigconf]{acmart}
\usepackage{url,hyperref}
\usepackage{textcomp,libertine}
\usepackage{amsmath,mathtools}
\usepackage{graphicx}

\usepackage{caption,wrapfig,subfigure,pdfpages,flushend}

\makeatletter

\newcommand{\Rmnum}[1]{\expandafter\@slowromancap\romannumeral #1@}
\makeatother

\makeatletter
\renewcommand\bibsection%
{
  \section*{\refname
    \@mkboth{\MakeUppercase{\refname}}{\MakeUppercase{\refname}}}
}
\makeatother

\acmYear{2017} 
\setcopyright{acmcopyright}
\acmConference[ACSAC 2017]{the 33rd Annual Computer Security Applications Conference}{December 4--8, 2017}{San Juan, PR, USA}

\begin{document}

\title{The Devil's in The Details: Placing Decoy Routers in the Internet}


\author{Devashish Gosain}
\affiliation{\institution{IIIT Delhi, India}}
\email{devashishg@iiitd.ac.in}

\author{Anshika Agarwal} 
\affiliation{\institution{IIIT Delhi, India}}
\email{anshika1448@iiitd.ac.in}

\author{Sambuddho Chakravarty}
\affiliation{\institution{IIIT Delhi, India}}
\email{sambuddho@iiitd.ac.in}

\author{H. B. Acharya}
\affiliation{\institution{Rochester Inst of Tech}}
\email{acharya@mail.rit.edu}

\acmISBN{978-1-4503-5345-8/17/12}
\acmDOI{10.1145/3134600.3134608}

\begin{abstract}
\emph{Decoy Routing}, the use of routers (rather than end hosts) as proxies, is a new direction in anti-censorship research. Decoy Routers (DRs), placed in Autonomous Systems, proxy traffic from users; so the adversary, e.g. a censorious government, attempts to avoid them. It is quite difficult to place DRs so the adversary cannot route around them -- for example, we need the cooperation of 850 ASes to contain China alone~\cite{houmansadr2014no}.

In this paper, we consider a different approach. We begin by noting that DRs need not intercept all the network paths from a country, just those leading to \emph{Overt Destinations}, \emph{i.e.} unfiltered websites hosted outside the country (usually popular ones, so that client traffic to the OD does not make the censor suspicious). Our first question is -- How many ASes are required for installing DRs to intercept a large fraction of paths from \emph{e.g.} China to the top-$n$ websites (as per Alexa)? How does this number grow with $n$ ? To our surprise, the same few ($\approx 30$) ASes intercept over $90\%$ of paths to the top $n$ sites worldwide, for $n = 10, 20 ... 200$ \emph{and also to other destinations}. Investigating further, we find that this result fits perfectly with the hierarchical model of the Internet \cite{subramanian2002characterizing}; our first contribution is to demonstrate with real paths that \emph{the number of ASes required for a world-wide DR framework is small} ($\approx 30$). Further, censor nations' attempts to filter traffic along the paths transiting these $30$ ASes will not only block their own citizens, but others residing in foreign ASes.

Our second contribution in this paper is to consider the \emph{details} of DR placement: not just in which ASes DRs should be placed to intercept traffic, but exactly where in each AS. We find that even with our small number of ASes, we still need a total of about $11,700$ DRs. We conclude that, even though a DR system involves far fewer ASes than previously thought, it is still a major undertaking. For example, the current routers cost over $10.3$ billion USD, so if Decoy Routing at line speed requires all-new hardware, the cost alone would make such a project unfeasible for most actors (but not for major nation states).
\end{abstract}


\maketitle






\section{Introduction}
\label{sec:introduction}

Anti-censorship systems such as proxies or \texttt{Tor}~\cite{dingledine2004tor} suffer from a double bind. To be useful, the entry point to the service must be discoverable to the user -- typically, the citizen of a censorious country. On the other hand, as soon as the entry point becomes common knowledge, it also comes to the attention of the censoring government, who shuts it down~\cite{winter2012great}. \emph{Decoy Routing}, a new anti-censorship paradigm\cite{karlin2011decoy,houmansadr2011cirripede,wustrow2011telex,wustrow2014tapdance,rebound15,bocovich2016slitheen}, attempts to disrupt this dynamic by using special routers as proxies, rather than end hosts. A DR lies on the path of traffic between the user inside a censorious country and an apparent (``overt'') destination; when it senses secret handshake data embedded in the user's packets, it intercepts the packets and re-sends the  message they carry to the real (``covert'') destination. Note that the DR, being outside the censorious country, can freely communicate with the covert destination -- and unlike an end-host proxy, cannot \textit{easily} be blacklisted. 
 
However, ``easy'' is a relative term. In their paper on ``Routing around Decoys''~\cite{schuchard2012routing}, Schuchard~\emph{et al.} propose that a sufficiently powerful adversary can simply route around ASes in the Internet where DRs are positioned. Houmansadr~\emph{et al.} ~\cite{houmansadr2014no} retort that such a move is extremely expensive, and in any case one could leave the adversary with no such option,~\emph{e.g.} by placing DRs in enough ASes to completely encircle a censorious country. They then follow up with a model~\cite{houmansadr2016game}, where they frame the problem of placing DRs, versus the problem of bypassing them, as an adversarial game. \textit{But the problem remains that the best known solutions still require the collaboration of several hundred ASes, in order to leave a \textbf{\emph{single}} well-connected country\footnote{A ``well-connected'' country does not just refer to major powers like China; even \emph{e.g.} Venezuela is well-connected in this regard.} with no choice but to route through one of them.} Further, such solutions require the computation of separate sets of ASes for each adversary nation \cite{houmansadr2014no,houmansadr2016game}.

\emph{Our first contribution in this paper is a new approach to the question of placing DRs.} In Decoy Routing, the router \emph{intercepts} messages, from the user inside a censorious nation, en route to an overt destination. What if, instead of trying to intercept \emph{all} the flows from a censorious country, we consider only the flows to the overt destinations? The overt destination is most likely a well-known site, often visited by citizens of the target country. [If not, it is very hard for users to discover; and when it is found, the sudden surge of traffic from users in China to some obscure website in e.g. Turkmenistan will itself make the censor suspicious.] 

As a first step, we started with the assumption that the overt destinations are popular sites (such as the Alexa top-$10$). We constructed a map of AS-level paths, connecting all ASes of the Internet to these, using the approach described by Gao \emph{et al.}~\cite{qiu2006cam04} (involving real BGP routing tables and inter-AS relationships~\cite{caida-inter-as-rels}). We then identified the \emph{``key'' ASes} -- those which appear most frequently on a large fraction of the paths. We find that $\approx 30$ ASes appear in more than $90\%$ of the paths to our target sites.  

Our approach in this first step is not general; clearly, the adversary could block access to the entire Alexa top-$10$, to prevent users' traffic from reaching the DRs. So our second step is a study of how the ``hardness'' of the problem -- finding the set of ASes that intercept
several AS paths -- varies as we change the set of possible overt destinations to the top-$10$, $20$, $30$, $50$, $100$, or $200$ web sites.

Interestingly, we found the \emph{same} set of $30$ ASes intercept
over $90\%$ of paths in all cases -- whether we consider paths leading to the top-$10$, $20$ ... or $200$. 
However, this result is easy to explain in hindsight. The Internet consists of ASes linked by peer-peer and provider-customer relationships; the ``top of the hierarchy''
or ``core'' consists of a few large multi-national ASes that peer with one another, and provide Internet access to most other ASes~\cite{subramanian2002characterizing,hu2005exploiting,magoni2005internet,orsini2014evolution}. Given such findings, and our experimental results with real paths, we come to a very powerful conclusion: only $30$ ASes, all in non-censorious countries, are sufficient for a DR infrastructure that intercepts more than $90\%$ of paths to important websites \emph{in general}. In such a case, besides the reduction in the number of ASes compared to current solutions (about $30$ times) \emph{this method needs to be run only once, rather than separately for each censorious country}. Our further experiments indicate that this is indeed the case -- the power of these ASes is not limited to the top-100 websites, they intercept over $90\%$ of paths for other destinations as well. For example, with nine case studies of censorious countries, we found that these key ASes also intercept over $90\%$ of the paths to $450$ websites that are popular across these nations and also hosted outside their respective network boundaries.

Our AS-level results suggest that censorious countries in the Internet are less able to ``route around decoys'' than previously thought. About 30 ASes -- $0.055\%$ of the world ASes -- intercept over $90\%$ of paths to popular websites, and in particular, $99\%$ of the paths originating from China. Furthermore, if censorious regimes choose to filter traffic along paths traversing the key ASes, they affect customers outside their network boundaries, and the extent of this ``collateral damage'' can be extremely high -- for example, over $92\%$ of all the network paths that traverse Chinese ASes originate beyond its network boundaries. (Details in section \ref{sec:discussion}.)

\textit{For our second contribution in this paper, we raise a new question}. DR placement is not limited to AS selection! A large AS has thousands of routers; where exactly in the AS should DRs be placed? In this first study that uses intra-AS mapping (\emph{viz.} \emph{Rocketfuel}~\cite{rocketfuel-paper}) to answer the aforementioned question, we find that while the number of ASes required for a world-wide DR framework is very small ($30$), we need to replace on average $400$ routers per AS with DRs. 

We conclude that, while a global DR system may involve only $30$ ASes, a practical one would still require placing over $11,700$ DRs in about $13$ different countries. In fact, the problem remains challenging even if we provide Decoy Routing to citizens of a single country: against a very weak adversary, Syria (contained by only $3$ ASes), a DR framework would involve $1,117$ routers. 
\emph{No existing DR architectures have been shown to process requests at line rates of network backbone routers\footnote{Which is of the order of several Tbps~\cite{chao2007high}}, nor has an implementation on existing high-speed routers been developed.} 
Unless we can deploy Decoy Routing on existing (or augmented) networking infrastructure, and can handle the high speeds, we will need to replace infrastructure at costs of over ten billion dollars (for example, for Level-3 Communications alone, \emph{i.e.} AS 3356 and AS 3549, the cost is $1.4$ billion USD at Parulkar and McKeown's~\cite{das2013rethinking}  estimate of $885,000$ USD/ router), \emph{plus} implementation, downtime, and debugging costs. 



\section{Background and Related Research}
\label{sec:back_rel}
This section presents the relevant background for our work, and a brief discussion of how it fits into the existing literature.

\subsection{Network Anti-censorship and Decoy Routing}

The general area of our work is the use of proxy servers to circumvent censorship. Popular anti-censorship solutions, such as Tor ~\cite{dingledine2004tor}\footnote{Onion routing was originally designed to ensure anonymity over the Internet, but as it tunnels encrypted messages through a distributed network of proxies, it is also suitable for evading censorship.}, are no longer powerful enough when the adversary is a sophisticated nation-state: there exist techniques to detect TLS flows carrying Tor ~\cite{chaabane2010digging,dharmapurikar2003deep}. More generally, traffic for most proxy based solutions can be detected and censored~\citep{ccs2012-skypemorph, ccs2012-stegotorus}, even if camouflaged~\cite{oakland2013-parrot}. 

\begin{figure}[h]
\centering
\includegraphics[scale=0.38]{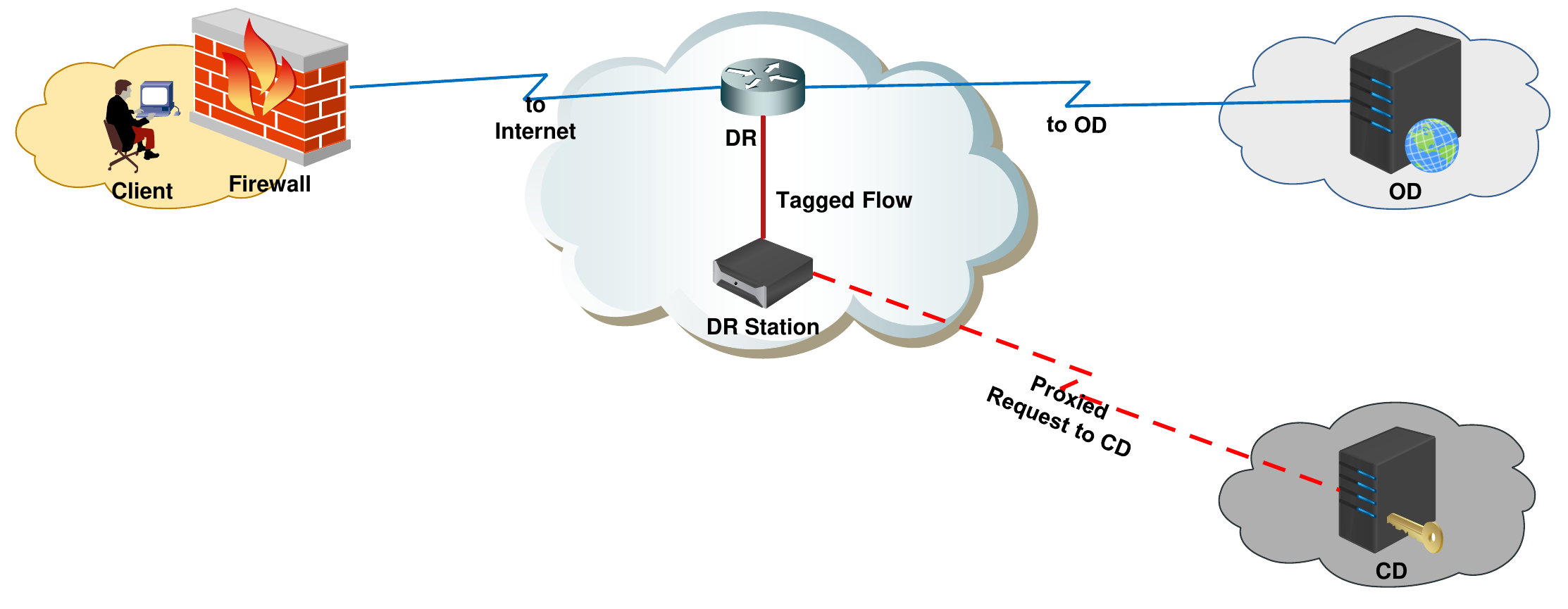}
\vspace{-0.5cm}\caption{\normalsize{Decoy Routing in Action: Clients in a censorious ISP bypass the filter by sending packets apparently addressed to a non-filtered overt destination (OD). En route, the packets traverse a DR, which sees the secret message; identifies them for special handling; decrypts them; and sends their payload to the real, covert destination (CD).\newline[Note: Current implementations \emph{cannot} perform Decoy Routing with just the router, they also require a DR station - typically a server - for the ``heavy lifting'', involving cryptographic operations]}}
\vspace{-0.5cm}
\label{fig:dr-basics}
\end{figure}

\emph{Decoy Routing}~\cite{karlin2011decoy} takes a new direction where proxying is performed by special network routers called \emph{Decoy Routers}. We sketch the basic mechanism in brief.
\begin{itemize}
\item The user of DR is hosted within a censorious ISP network, but wishes to communicate with network destinations censored by its ISP. To achieve this, it sends packets addressed to an innocuous-looking website, known as the \emph{overt destination}. (The packets are encrypted using TLS, so the ISP cannot see the contents, and the header shows that they are meant for an unfiltered destination.) 

\item These innocuous-appearing packets, allowed out of the censoring ISP, carry a small, steganographic message, usually encoded in the protocol headers.

\item On their way to the overt destination, if the packets pass through a DR, the steganographic message acts as a secret handshake. Instead of forwarding them, the DR decrypts their payload (the key, the TLS shared secret, is also sent as part of the secret message); establishes a new connection to the filtered site - the true, \emph{covert destination}; and sends the payload to this covert destination. 

\end{itemize} Thus, a DR acts as a proxy, covertly communicating with a blocked site on behalf of the user. This procedure, \emph{end-to-middle (E2M)} censorship circumvention, is shown in Figure~\ref{fig:dr-basics}. Actual implementations of Decoy Routing -- Telex~\cite{wustrow2011telex}, Cirripede~\cite{houmansadr2011cirripede}, TapDance~\cite{wustrow2014tapdance}, Rebound~\cite{rebound15} and Slitheen \cite{bocovich2016slitheen} -- have different features (message replay protection, tolerance of asymmetry in routing, inline blocking of traffic to/from overt destination, implementation of secret handshake, \emph{etc.}), but share the basic design outlined above. This design decision stems from the realization that it is much harder for the censor to prevent the packets passing through a router, than it is to block an end host. But \emph{how} hard it really is for the censor to circumvent DRs, and \emph{where} the routers should be placed, is an active research question, as we discuss in the next sub-section.

\subsection{On The Placement of DRs}

Where should DRs be placed in the Internet? This question was first raised by the Cirripede project \cite{houmansadr2011cirripede}, whose authors claim that (against an adversary who is ignorant of Decoy Routing), placing DRs in just two tier-1 ASes is sufficient to serve all clients worldwide. 

The next major step, by Schuchard \emph{et al.} \cite{schuchard2012routing}, is to suggest that a powerful adversary such as China will eventually figure out which ASes have DRs in them, and simply redirect its traffic to avoid these ASes -- the Routing Around Decoys (RAD) attack. Mapping the Internet at AS level (ASes and their connections), the authors show that censor countries (China, Iran, \emph{etc.}) have connections to many ASes, and thus enough alternative paths to route around a particular AS. Avoiding the top $100$ ASes (by degree in the CAIDA connectivity graph) would disconnect China from only $2.3\%$ of web destinations. 

Houmansadr \emph{et al.}~\cite{houmansadr2014no} counter that, once we consider actual routing -- with directional business relationships between ASes, rather than just graph connectivity -- the RAD attack is too costly to be feasible. They also question Schuchard~\emph{et al.}'s assumption that DRs may be placed in randomly-chosen ASes. $86.2\%$ of the ASes on the Internet are \emph{origin} ASes (\emph{i.e.} they do not transport traffic of other ASes); a random placement mostly chooses such ASes, and it is possible to do much better if the ASes are chosen strategically. The authors propose two ways to do this: 
\begin{enumerate}
\item \emph{Sorted placement}. ASes that appear most frequently in the \emph{adversary's} routing tables. 
\item \emph{Strategic random}. ASes chosen randomly, but only among those ASes that have a large enough \emph{customer-cone}. \footnote{Customer cone refers to customers, customers of customers, etc. In other words, a selected AS must be a significant provider to other ASes.}
\end{enumerate} 
But while this approach is better than random, it still computes a separate, large set of ASes for every adversary ($858$ ASes for China, $835$ for Venezuela, etc.). 

Further, Kim~\emph{et al.} \cite{rebound15} also
suggest a graph theoretic approach to solve the problem, involving
hypothetical network graphs, without however considering how network
routes are determined by inter-AS relationships~\cite{gao2001inferring}.

The first contribution of this paper is a new approach for placing DRs: we select the ASes that appear most frequently in paths from \textbf{all} ASes to popular websites (potential Overt Destinations), as candidates for placing DRs -- as estimated from actual routing tables.  

In our experiments, we target multiple sets of target websites (globally popular ones, those popular in selected censorious nations, etc.), and consistently find that \emph{the same set of key ASes cover the vast majority ($>90\%$) of routes to the target}. (Genuine Internet BGP routes, collected by Routeviews \cite{routeviews}, not a simulation.) We suggest that these ``heavy-hitter'' ASes are very likely the current ``core'' of the Internet, as first found by Rexford \emph{et al} \cite{subramanian2002characterizing}, and are good candidates for installing DRs as they intercept the vast majority of flows to any destination. 

Our results indicate that \emph{30 ASes suffice to provide Decoy Routing worldwide} (in comparison, the state of the art \cite{houmansadr2014no} requires over 850 ASes to contain a single adversary country, China).

A possible objection to our method is that several of the ``heavy-hitter'' ASes of the Internet~\cite{cesareo2012optimizing} may themselves be adversaries, as they are hosted in censorious nations ~\cite{lcn2017}. However, in this paper, we show that only a handful of ASes ($\approx$ 30) are needed to host Decoy Routers, \emph{even when we restrict ourselves from using ASes in adversary countries such as Russia or China.} 


Moreover, we correct some incorrect assumptions made by the earlier authors. Houmansadr \cite{houmansadr2014no} use customer-cone size as a metric to choose ASes, assuming that it is a good predictor of how many flows they carry; we explain in Section~\ref{sec:discussion} (and in the appendix) why it is \textbf{not}.

Our second, and more important, contribution is to demonstrate that \emph{even though the number of ASes needed for a DR infrastructure is small, the actual number of routers that are to be replaced with DRs is large.} We map ASes at the router-level, using Rocketfuel~\cite{rocketfuel-paper}, and identify the specific network elements that potentially need to be replaced by DRs; on average, for each AS we need to deploy several hundred DRs. We suggest that the cost of such ``major surgery'' effectively removes the possibility that ASes would operate such a project \emph{pro bono}, and raises the question of how such an infrastructure may be economically feasible ~\cite{houmansadr2016game}.





\subsection{Mapping the Internet}

Our work depends on finding the paths to a particular destination taken by Internet traffic. In this sub-section, we give a short introduction to Internet mapping, and explain our method of mapping.

The Internet consists of routers and hosts, organized into networks called Autonomous Systems (ASes). These networks operate independently, but collaborate to route traffic among themselves. ASes can be customers, peers, or providers to other ASes; besides a physical connection, there must be an acceptable business relationship between two ASes, before they route traffic through each other.\footnote{A customer AS routes traffic through its providers; but providers do not route ``through'' traffic through their customers. The only traffic a provider sends a customer, is meant for that customer, or \emph{its} customers, and so on. }


Mapping the Internet involves two tasks -- Finding inter-AS connections (and relationships) and mapping routers and hosts (and their connections, inside ASes). 

\noindent \textbf{AS-level mapping.:}
Projects such as CAIDA Ark~\cite{caida-ark} and \emph{iPlane}~\cite{madhyastha2006iplane} map Internet routes with \texttt{traceroute}. Traceroute returns router-level paths from a source to a destination, hop-by-hop; the map is built by running traceroute from distributed volunteer nodes to various $\//24$ prefixes. This data is consolidated into a graph where the nodes represent ASes, and edges represent links between them. 

Such approaches are generally limited by the network locations and availability of the volunteer nodes; they may not provide the AS-level path between any two randomly chosen ASes, and even where they do, they may be inaccurate. 

In our research we used the approach of Gao and Qiu \cite{qiu2006cam04}, that uses RIBs collected from the Routeviews project \cite{routeviews} and ``stitches together'' known links, thus constructing paths to our target sites from every AS in the Internet. This approach has been used in the past by others~\cite{pets2016-denasa,DBLP:conf/ccs/EdmanS09}. Details are given in Appendix \ref{sec:app_gao}.

\noindent \textbf{Router-level mapping:}
A large AS, such as an ISP, generally has several thousand routers. In theory, it is possible to repeat our approach for inter-AS mapping (where we use BGP information), and map the internal structure of ASes using their SNMP Management Information Bases (MIBs) \cite{pandey2009snmp}. However, we have no access to this data. Instead, we mapped the routers in ASes of interest using the Rocketfuel approach \cite{rocketfuel-paper} (this involves running \texttt{traceroute} probes from looking glass servers~\cite{traceroute-looking-glass} to prefixes inside a chosen AS). Thereafter, \emph{IP aliases}\footnote{Different interfaces of the same router, with different IP addresses, are called IP aliases} are resolved using \texttt{Midar}~\cite{midar-caida}.
 

\section{Motivation}
\label{sec:motivation}

The problem in this paper is to determine where in the Internet we should place DRs, in order to intercept large fraction
of network paths. 
The current state of the art ~\cite{houmansadr2014no} chooses ASes which are strongly linked with each target country (therefore intercepting much of their traffic), and whose customer cone size exceeds some threshold. However, this approach has the following limitations:
\begin{enumerate} 
\item New ASes must be identified for each adversary nation.
\item This set of ASes is quite large. (over $850$ ASes for China,  nearly $850$ for Venezuela, etc.)
\item Customer-cone size does not seem to be an effective metric for choosing ASes that appear frequently in real routes (candidates for DR placement)\footnote{We mention the reasons in Section \ref{sec:discussion}. Details are provided in the Appendix.}
\item A large AS has thousands of routers, spread across several countries. Current methods identify the ASes to place DRs in - but not \emph{where} in the AS they should be placed.
\end{enumerate}

In order to address these limitations, we construct a map of the Internet, and select the ASes that occur most frequently in our paths (estimated using real BGP routing tables), instead of any other metric. Next, we map these ASes to identify their key routers; this allows us to estimate the number of DRs we need to be able to intercept a large fraction of Internet traffic. 

\section{Methods: Data Collection and Algorithm}
\label{sec:approach}

This section presents our algorithm for identifying key ASes in the Internet, and key routers in these ASes. Our focus in this section is on finding ASes and routers that intercept the paths from all ASes to important destinations (Alexa top-10, top-20 etc.) We also describe how we verify that our results are more general, \emph{i.e.} that our key ASes and routers also intercept paths to other destinations besides the top-$n$ website. This is covered in more detail in Section \ref{sec:discussion}.
Our network mapping process consists of two phases.\footnote{Our original plan was to map the entire Internet at the router level, and identify the key routers directly. Unfortunately, no existing techniques scale to mapping the Internet accurately at such fine granularity.}
\begin{itemize}
\item First, we build an AS-level Internet map, consisting of paths connecting popular websites and all the ASes of the Internet. We identify ASes that appear most frequently in those paths as key ASes (for hosting DRs).

\item In the second phase, we estimate the router -- level topology of key ASes to identify key routers -- the actual routers inside the ASes that transport the majority of traffic. 
\end{itemize}

\subsection{Generating AS level maps}
\label{sec:approach:mapping}
For the first phase of network mapping, we used the approach presented by Gao \emph{et al.}~\cite{qiu2006cam04}. AS paths are collected from BGP paths at Internet Exchange Points (IXes)~\cite{routeviews}. These tables, however, do not contain paths originating at every AS; Gao \emph{et al.}'s approach infers paths originating from every AS, using the existing BGP paths. ASes are appended to existing paths by selecting those that most frequently appear adjacent to ASes on the said BGP paths, without invalidating the path's \emph{valley-free} property\footnote{The AS-level path between two hosts on the Internet is said to follow a ``valley free'' path, as the path first rises - an AS, then its provider, then a provider of the provider, etc.; peaks - or plateaus, as it crosses through several peering links - and then descends, through provider-to-customer links, until it reaches the destination.
There must be no no provider-to-customer links between two customer-to-provider links (``valleys'')}. The aim is to build paths connecting every AS in the Internet to a given IP prefix. Details of the AS mapping algorithm are presented in Appendix~\ref{sec:app_gao}. 
For our analysis, we used snapshots of BGP RIBs collected from $15$ vantage points~\cite{routeviews}. Our original approach involved choosing the top-$10$ most popular sites, finding the paths from all ASes to their corresponding prefixes, and identifying the \emph{most frequently appearing ASes} on these paths. 

As presented in Section \ref{sec:eval_data}, we found a small set of ASes that appear in more than $90\%$ of the paths to these popular destinations from all ASes. We then increased the number of popular destinations -- top-$10$, $30$, $50$, upto $100$ -- and estimated the paths to the corresponding prefixes from all ASes. At each step we identified the set of ASes which appear most frequently in the paths. As we show in Section \ref{sec:eval_data}, the rough set of ASes remained almost unchanged as we varied the number of destinations.

These results suggest that the ASes  identified were likely ``heavy-hitter'' ASes of the Internet, potentially suitable candidates for DRs placement, as they may intercept a large fraction of network paths, originating at ASes around the world. As a caveat, we know that the Internet has a hierarchical structure, rooted at a few ``core ASes'' which peer with one another and intercept a large fraction of network routes~\cite{subramanian2002characterizing}.  But to test this claim, we had to answer two questions -- (a) Was it necessary to select exactly, and all, these ASes for placing DRs? Some of these ASes were in censorious countries. (b) How could we validate that our observations, that a great majority of paths were intercepted by these ASes, were not limited to the target OD sites we studied?

In order to answer the first question, we investigated the impact of replacing our key ASes in Russia, China, \emph{etc.}, with the next best choice: ASes ranked 31-50 by path frequency, but in non-censorious countries. We found the path coverage remained over $90\%$.

To answer the second question, we took the key ASes computed for the top-$100$ sites (say, Set-A). Next, taking sites ranked $101$-$200$ on Alexa (Set-B), we computed the paths to these sites for all ASes. We discovered that the key ASes, \emph{computed using the paths for Set-A}, continue to intercept over $90\%$ paths for Set-B.

Finally, we also computed the paths corresponding to the $50$ most popular websites in each of nine different censorious nations (say, Set-C). The same key ASes also intercepted over $90\%$ of the AS-level paths to destinations in Set-C.

Several months after our initial route collection, we repeated our experiments, and found the same set of key ASes intercepting over $90\%$ of the paths to Set-A, Set-B, and Set-C.

Our approach differs from what was proposed previously ~\cite{houmansadr2014no,schuchard2012routing}. The authors either chose Tier--1 ASes, or those that had large customer-cone sizes. We show in Section~\ref{sec:discussion}, and in the Appendix, that customer-cone size is poorly correlated to the number of network paths that traverse an AS (path frequency)--\emph{the latter being a better metric to select candidate ASes for DR placement.}

We note that Gao~\emph{et al.}'s algorithm generates the \emph{request} paths (connecting all ASes \emph{to} selected IP-prefixes), and not the \emph{reply} paths (\emph{from} the prefixes to the ASes). It is natural to ask whether asymmetry in routing might impact the strategy for placing DRs. However, the latest DR architectures, such as TapDance ~\cite{wustrow2014tapdance}, are \emph{agnostic} to path symmetricity\footnote{Responses from the overt destination are suppressed by manipulating http protocol states, without requiring the DR's intervention.}. 
This greatly simplifies the Decoy Router placement problem: we only need to place a DR on the path from the user to the OD, and not necessarily on the return path.

\vspace{-1mm}
\subsection{Creating router level maps} 

After identifying key ASes in the Internet, as above, we were still left with the problem of \emph{where} in the AS to put DRs. An AS involves a complex topology of routers and hosts; even the AS administrator, who knows the internal topology, may not know how frequently a router appears in actual network paths. When approaching AS admins to ask them to implement Decoy Routing, it is helpful to estimate how many (and which) routers they will need to replace. We therefore identified the actual routers that transport most of the ASes' traffic, using Rocketfuel~\cite{rocketfuel-paper} as follows: 








\begin{itemize}

\item For each chosen AS, we identified the prefixes it advertises (from \texttt{cidr-report.org}). From $390$ planetlab nodes, we targeted \texttt{Traceroute} probes to three representative IP addresses, corresponding to each prefix; we thus obtain router-level paths terminating at these prefixes. To capture paths transiting the said AS, we also ran \texttt{traceroute} probes  targeted towards IP prefixes in its neighboring ASes.

\item Using \texttt{Whois} ~\cite{cymru-ipwhois}, we inspected each \texttt{traceroute} paths to identify the first and last IP address belonging to the target AS. We denote these as the \emph{edge routers} of an AS (as opposed to \emph{core routers}, i.e. the internal routers of the AS). We trim the traces down to the part between these edge routers, \emph{i.e.} inside the AS. 

\item The router IPs (belonging to the target AS), discovered through the above process, suffer from problems such as aliasing \cite{network-map-survey}, so we resolved these aliases using the state-of-the-art alias resolution tool \texttt{Midar}~\cite{midar-caida}. 

\item Finally, from the \texttt{traceroute} results we identified a minimum number of routers which cumulatively intercept over $90\%$ of \texttt{traceroute} paths. When possible, we selected edge routers (as the edge routers cover $100\%$ of paths through the AS). But in cases where some heavy-hitter edge and core routers intercept over $90\%$ of paths, \emph{and this set is smaller than the set of edge routers}, we selected those (the former) instead.
\end{itemize}

\section{Data and Evaluations}
\label{sec:eval_data}

\begin{figure}[h!]
\centering
\normalsize
\vspace{-0.5cm}
\includegraphics[width=9cm]{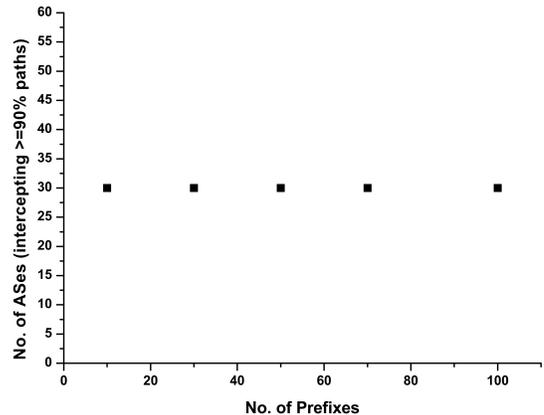}
\vspace{-1.35cm}
\caption{\normalsize{ASes needed to capture $90\%$ of traffic paths to different sets of overt destinations (popular websites).}}
\vspace{-0.5cm}
\label{fig:stable-set}
\end{figure} 

\subsection{Identification of Key ASes}

As described in the previous section, we began by selecting a small set of globally popular websites (Alexa top-$10$), computed the AS-level paths to them, and identified ASes which appeared most frequently in these paths. We recomputed such paths by increasing the number of popular websites -- top-$30$, $50$, $70$ and $100$. As figure~\ref{fig:stable-set} shows, the same number of roughly $30$ ASes intercepted over $90\%$ of the paths to these sites.

\begin{figure}
\centering
\vspace{-0.5cm}
\includegraphics[scale=0.32]{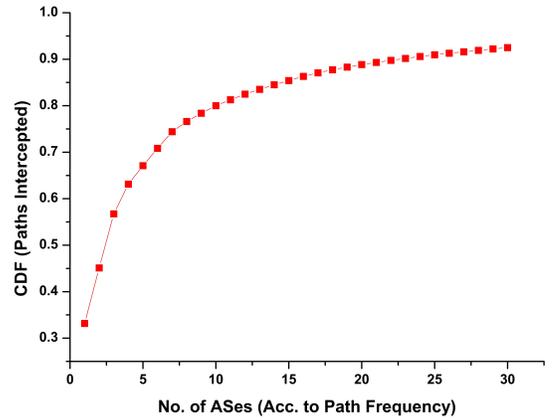}
\vspace{-0.75cm}
\caption{\normalsize{CDF: ASes and the fraction of paths they intercept. 
(CDFs are for paths to Alexa top-100 websites).}}
\label{fig:asfreqrankcu100}
\end{figure}

For instance, in Figure~\ref{fig:asfreqrankcu100}, we show the CDF of the $4,497,547$ paths connecting the Alexa top-$100$ sites to all ASes, and their interception by our top-30 ASes. The X-axis represents the top-$30$ ASes ranked by their path frequencies; the Y-axis represents the actual fraction of paths. The highest-ranked AS, AS3356 (Level 3 Communications), intercepts $1,492,079$ paths ($33.2\%$ of all paths). The top $2$ ASes, AS3356 and AS174 (Cogent Communications), intercept $2,028,831 (= 1,492,079 + 536,752)$ unique IP-prefix-to-AS paths, ($45.1\%$ of all  paths). The top $30$-ASes by path frequency together intercept $92.4\%$ of all paths. 

However, there is a major problem with considering the top-$30$ ASes as the preferred location for placing Decoy Routers. In Table 1, which presents the corresponding ASes, their hosting country, and their ranks based on path frequency ($P_{freq}$) and customer-cone size ($C_{size}$), we highlight this problem:  a substantial fraction of these ASes lie in countries known to censor Internet traffic, such as Russia and China~\footnote{As per censorship ratings by Freedom House Report~\cite{fh-report} and the ONI~\cite{oni-url}.}. So the question arises whether we can find acceptable alternatives.

\begin{figure}[t!]
\vspace{-0.5cm}
\centering
\includegraphics[scale=0.36]{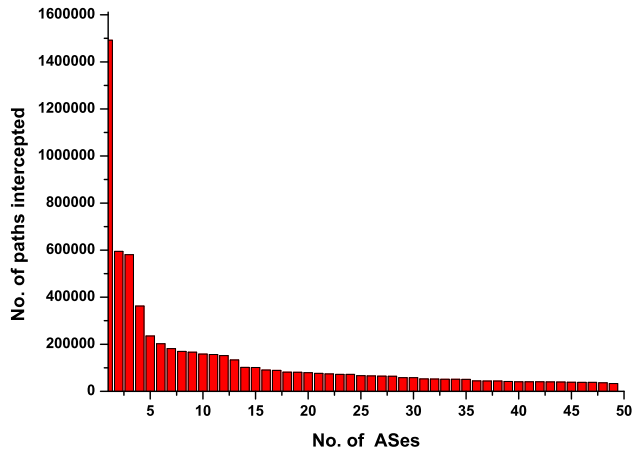}
\vspace{-1.25cm}
\caption{\normalsize{No. of paths intercepted by each of the top-50 ASes (sorted by path frequency).}}
\label{fig:as-freq-rank}
\vspace{-0.5cm}
\end{figure}

As we see in Figure \ref{fig:as-freq-rank}, while the number of \emph{unique} paths intercepted falls off rapidly, the \emph{total} paths intercepted, \emph{including overlaps}, does not. This observation emboldens us to suggest that ASes ranked $31-50$ are comparable to those ranked $11-30$ in terms of the paths they intersect. Accordingly, from the ASes ranked  between $31-50$, we selected $9$ new ASes headquartered in non-censorious regimes as replacements for the (likely) hostile ASes in Table 1. These ASes are presented in Table~\ref{tab:tab_non_censor}.

\begin{figure}
\vspace{-0.5cm}
\centering
\includegraphics[scale=0.3]{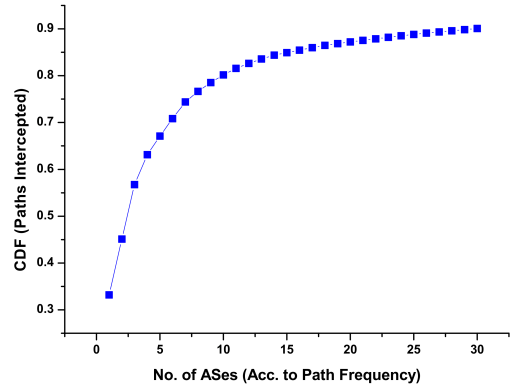}
\vspace{-0.75cm}
\caption{\normalsize{CDF of ASes (hosted in non-censorious ASes) according to fraction of paths that they intercept.}}
\label{fig:as-freq-rank_cu_without_censor}
\end{figure}

Figure \ref{fig:as-freq-rank_cu_without_censor} presents the proportion of paths to Alexa top-100 websites covered by our new set of key ASes  (\emph{i.e.} redefining \emph{key AS} to exclude ASes in censorious nations). We see by comparing with Figure~\ref{fig:asfreqrankcu100}, that \emph{the chosen set of key ASes are roughly as effective at intercepting traffic as the top-30 ASes}. (To be exact, they intercept $90.2\%$ of the AS-to-prefix paths, compared to $92.4\%$ for the top-$30$.) In fact, they do so consistently, for all target prefixes in our tests.

\begin{table}
\small
\centering
\scalebox{0.85}{
\begin{tabular}{|l|l|l|l|}
\hline
\textbf{ASN} & \textbf{Country} & \textbf{Rank (P$_{freq}$)} & \textbf{Rank (C$_{score}$)}\\\hline
3356 & US & 1 & 1 \\\hline
174 & US & 2 & 2 \\\hline
2914 & US & 3 & 5 \\\hline
1299 & SE & 4 & 4 \\\hline
3257 & DE & 5 & 3 \\\hline
6939 & US & 6 & 13 \\\hline
6461 & US & 7 & 8 \\\hline
6453 & US & 8 & 52 \\\hline
7018 & US & 9 & 17 \\\hline
10310 & US & 10 & 6 \\\hline
\color{red} 4134 & \color{red} CN & \color{red} 11 & \color{red} 10 \\\hline
3549 & US & 12 & 79 \\\hline
\color{red} 4837 & \color{red} CN & \color{red} 13 & \color{red} 85 \\\hline
209 & US & 14 & 19 \\\hline
9002 & UA & 15 & 97 \\\hline
\color{red} 6762 & \color{red} IT & \color{red} 16 & \color{red} 7 \\\hline
\color{red} 8359 & \color{red} RU & \color{red} 17 & \color{red} 22 \\\hline
2828 & US & 18 & 30 \\\hline
\color{red} 20485 & \color{red} RU & \color{red} 19 & \color{red} 21 \\\hline
16509 & US & 20 & 9 \\\hline
\color{red} 9498 & \color{red} IN & \color{red} 21 & \color{red} 18 \\\hline
4323 & US & 22 & 16 \\\hline
\color{red} 3216 & \color{red} RU & \color{red} 23 & \color{red} 99 \\\hline
2497 & JP & 24 & 15 \\\hline
701 & US & 25 & 12 \\\hline
12956 & ES & 26 & 65 \\\hline
37100 & MU & 27 & 23 \\\hline
\color{red} 4826 & \color{red} AU & \color{red} 28 & \color{red} 26  \\\hline
\color{red} 12389 & \color{red} RU & \color{red} 29 & \color{red} 67 \\\hline
1335 & US & 30 & 92 \\\hline
\end{tabular} }
\caption{\normalsize {Top 30 ASes that intercept more than $90\%$ of paths. (ASes headquartered in censor nations are highlighted.)}}
\label{tab:Top30AS}
\vspace{-0.5cm}
\end{table}
\begin{table}
\small
\centering
\scalebox{0.85}{
 \begin{tabular}{|l|l|l|l|} 
 \hline
\textbf{ASN} & \textbf{Country} & \textbf{Rank (P$_{freq}$)} & \textbf{Rank (C$_{score}$)}\\\hline
\color{blue} 13030 & \color{blue} SW & \color{blue} 31 & \color{blue} 84 \\\hline
\color{blue} 1273 & \color{blue} UK & \color{blue} 32 & \color{blue} 83 \\\hline
\color{blue} 16735 & \color{blue} BZ & \color{blue} 33 & \color{blue} 98 \\\hline
\color{blue} 6830 & \color{blue} EU & \color{blue} 34 & \color{blue} 91\\\hline
\color{blue} 18881 & \color{blue} BZ & \color{blue} 35 & \color{blue} 95 \\\hline
\color{blue} 3491 & \color{blue} US & \color{blue} 36 & \color{blue} 42\\\hline
\color{blue} 10026 & \color{blue} HK & \color{blue} 37 & \color{blue} 87 \\\hline
\color{blue} 32787 & \color{blue} US & \color{blue} 39 & \color{blue} 93 \\\hline
\color{blue} 1239 & \color{blue} US & \color{blue} 46 & \color{blue} 45\\\hline
\end{tabular}
}
\vspace{0.15cm}
\caption{\normalsize{ASes hosted in non-censorious nations ranked by path frequency (ranks \textgreater$30$ and \textless$50$)}}
\vspace{-0.5cm}
\label{tab:tab_non_censor}
\end{table}

\subsection{Identifying important routers inside key ASes}

The second part of our research involves identifying the important routers inside key ASes. As described in Section~\ref{sec:approach:mapping}, we used \texttt{Traceroute} to probe IPs in each prefix advertised by the key ASes. From these traces we determined the candidate routers that may be replaced 
with DRs. 
 
We originally chose to n\"{a}ively replace edge routers with DRs, as these intercept all traffic entering and leaving an AS. However, we found that in many cases the total number of edge routers is significantly greater than the number of ``heavy-hitter'' routers -- a set of edge and core routers that collectively appear in more than $90\%$ of the \texttt{traceroute} paths for the AS. We therefore updated our approach. For each AS, we now find both sets (edge routers and heavy-hitter routers), and select the smaller set as the \emph{key} routers, \emph{i.e.} the candidates for being replaced with DRs. For example, for AS $4134$, we need only $179$ heavy-hitter routers (including both edge and core routers) to capture over $90\%$ of the paths, but $749$ edge routers to intercept $100\%$ paths, while for AS209 (Quest Communications), we choose the edge router set -- about $1662$ routers. We present our results in Table~\ref{tab:tab_Inside_AS}.

\begin{table}[b!]
\vspace{-0.32cm}
\small
\centering
\scalebox{0.9}{
 \begin{tabular}{|l|l|l|l|l|} 
 \hline
 \textbf{ASN} & \textbf{\# of}           &   \textbf{\# of}       &   \textbf{\# of}        & \textbf{\# of}\\
              & \textbf{Edge}           &   \textbf{Core}       &   \textbf{Heavy}       & \textbf{DR's}\\
              & \textbf{Routers}        &  	\textbf{Routers}    &   \textbf{Hitter}      & \textbf{Required}\\
              & \textbf{(E) }           &   \textbf{(C) }       &   \textbf{Routers} &    \textit{min}\textbf{(E, H)} \\ 
              &                         &                       &    \textbf{(H)}    &  \\\hline
              
3356 & 707 & 303 & 576 & \textbf{576}\\\hline
174 & 165 & 1572 & 288 & \textbf{165} \\\hline
2914 & 134 & 2061 & 534 & \textbf{134}\\\hline
1299 & 493  & 1989 & 517 & \textbf{493}\\\hline
3257 &  762  & 2316 & 1483 & \textbf{762}\\\hline
6939&   169&	554&	103 & \textbf{103}\\\hline
6461&   105&	850&	45& \textbf{45}\\\hline
6453&   223&	896&	210 & \textbf{210}\\\hline
7018&   359&	6003& 107 & \textbf{107}\\\hline
10310&  161& 156& 106 & \textbf{106}\\\hline
\color{red} 4134& \color{red} 749 & \color{red} 10078 & \color{red} 177 & \color{red} \textbf{177}\\\hline
3549&	943& 6227&5579 & \textbf{943}\\\hline
\color{red} 4837& \color{red} 1031 & \color{red} 7350 & \color{red} 2538 & \color{red} \textbf{1031}\\\hline
209&	1662&10842&8687 & \textbf{1662}\\\hline
9002&	30&	47&	40 & \textbf{30}\\\hline
\color{red} 6762& \color{red} 154& \color{red} 333& \color{red} 238 & \color{red} \textbf{154}\\\hline
\color{red} 8359& \color{red} 25& \color{red} 320& \color{red} 13 & \color{red} \textbf{13}\\\hline
2828&	116&	1049& 636 & \textbf{116}\\\hline
\color{red} 20485&	\color{red} 506& \color{red} 206& \color{red} 193 & \color{red} \textbf{193}\\\hline
16509&	1244&	5311& 4644 & \textbf{1244}\\\hline
\color{red} 9498 & \color{red} 320 & \color{red} 199 & \color{red} 269 & \color{red} \textbf{269}\\\hline
4323 & 668 & 2548 & 2695 & \textbf{668}\\\hline
\color{red} 3216& \color{red} 305& \color{red} 1981& \color{red} 1769 & \color{red} \textbf{305}\\\hline
2497 &	187 & 1078 &	133 & \textbf{133}\\\hline
701	& 1770 & 4417& 2975 & \textbf{1770}\\\hline
12956 & 482 & 734& 681 & \textbf{482}\\\hline
37100 & 14 & 72& 59 & \textbf{14}\\\hline
\color{red} 4826& \color{red} 43& \color{red} 381& \color{red} 30 & \color{red} \textbf{30}\\\hline
\color{red} 12389&	\color{red} 322& \color{red} 2625& \color{red} 1898 & \color{red} \textbf{322}\\\hline
\end{tabular}
}
\vspace{0.25cm}
\caption{\normalsize{
Edge routers, core routers, heavy-hitter routers and the routers required for replacement with 
DRs. Applying our router selection strategy, for \emph{e.g.} for AS3356 --
edge routers: $707$ core routers: $303$.
Routers (both edge and core) covering $90\%$ of the paths: $576$. We thus select the latter. Total routers required for all the 30 ASes (headquartered in censorious and non-censorious nations) : $12,257$.}}
\label{tab:tab_Inside_AS}
\vspace{-0.5cm}
\end{table}


As mentioned previously, several of these ASes are hosted in censorious regimes, and so we identified the number of routers to be replaced with DRs in non-censorious 
countries. While the results
presented in table~\ref{tab:tab_Inside_AS}
represents the number of routers
to be replaced for ASes presented in Table $1$, Table~\ref{tab:tab_Inside_AS_noncensor_1}
represent the number of routers for ASes
in non-censorious nations presented in Table~\ref{tab:tab_non_censor}.

The total number of routers that may be
replaced across ASes in non-censorious ASes
is $11,709$.


\begin{table}[b!]
\small
\centering
\scalebox{0.9}{

\begin{tabular}{|l|l|l|l|l|} 
 \hline
 \textbf{ASN} & \textbf{\# of}           &   \textbf{\# of}       &   \textbf{\# of}        & \textbf{\# of}\\
              & \textbf{Edge}           &   \textbf{Core}       &   \textbf{Heavy}       & \textbf{DR's}\\
              & \textbf{Routers}        &  	\textbf{Routers}    &   \textbf{Hitter}      & \textbf{Required}\\
              & \textbf{(E) }           &   \textbf{(C) }       &   \textbf{Routers} &    \textit{min}\textbf{(E,H)} \\ 
              &                         &                       &    \textbf{(H)}    &  \\\hline

\color{blue} 13030 & \color{blue} 58 & \color{blue} 302 & \color{blue} 38 & \color{blue} \textbf{38}\\\hline
\color{blue} 1273 & \color{blue} 156 & \color{blue} 1106 & \color{blue} 693 & \color{blue} \textbf{156}\\\hline
\color{blue} 16735 & \color{blue} 12 & \color{blue} 43 & \color{blue} 37 & \color{blue} \textbf{12}\\\hline
\color{blue} 6830 & \color{blue} 216 & \color{blue} 4048 & \color{blue} 1654 & \color{blue} \textbf{216}  \\\hline
\color{blue} 18881 & \color{blue} 338 & \color{blue} 3893 &  \color{blue} 431 & \color{blue} \textbf{338} \\\hline
\color{blue} 3491 & \color{blue} 698 & \color{blue} 1139 &  \color{blue} 955 & \color{blue} \textbf{698} \\\hline
\color{blue} 10026 & \color{blue} 170 & \color{blue} 765 &  \color{blue} 346 & \color{blue} \textbf{170}\\\hline
\color{blue} 32787 & \color{blue} 46 & \color{blue} 571 &  \color{blue} 456 & \color{blue} \textbf{46}\\\hline
\color{blue} 1239 & \color{blue} 242 & \color{blue} 1221 & \color{blue} 910 & \color{blue} \textbf{242}\\\hline
\end{tabular}
}
\vspace{0.25cm}
\caption{\normalsize{Edge routers, core routers, heavy-hitters, and required DRs, for our ``replacement'' key ASes (from Table ~\ref{tab:tab_non_censor}).}}
\label{tab:tab_Inside_AS_noncensor_1}
\vspace{-0.75cm}
\end{table}
The $11,709$ key routers, across 30 key ASes that together intercept greater than $90\%$ of network paths, together represent a formidable infrastructure, with equipment costs of $10.3$ billion dollars. Converting these routers to DRs would involve massive system implementation, testing, deployment and related costs.


\noindent \textbf{Implementation Details:}
Our AS-level map uses BGP Routing Information Base (RIB) data, which we obtain from $15$ Internet Exchange points through routeviews ~\cite{routeviews}, and AS relationship data from CAIDA~\cite{caida-inter-as-rels}. The map was constructed using virtual machines with a total of $10$ CPU cores (x64) and $24$ GB RAM, running Ubuntu Linux (14.04LTS server). Our  multi-threaded implementation of Gao's \cite{qiu2006cam04} algorithm took $\approx$ 3-4 hrs. to compute paths to $10$ prefixes. 

To identify key routers in an AS, we ran \texttt{traceroute} probes from $390$ \texttt{planetlab} machines to three random IP's in each prefix advertised by the AS. Depending upon the number of prefixes advertised, and network latency, it took approximately $18-36$ hours to probe an AS, and $5-8$ hours for alias resolution.

\section{Data Analysis and Discussion}
\label{sec:discussion}

Our method for placement of DRs, as presented in this paper, has several major advantages: 
\begin{enumerate}
\item The placement of DRs is \emph{global}, and needs to be run \emph{only once} to provide a small list of ASes that cover paths from all adversaries. (Existing approaches~\cite{houmansadr2014no} require fresh candidate ASes to place DRs for each adversary nation.)

\item The ASes selected are located far away from the adversary nations, and thus outside their geo-political and economic sphere of control. This makes it more difficult to bring pressure to bear on them. 

\item The selected ASes lie on a very large fraction of paths. It is therefore hard for RAD adversaries ~\cite{schuchard2012routing} to bypass them without risking disconnection from all or most of the Internet. (This might inadvertently also disconnect foreign customers, as we see in sub-section 6.2 .) 
\end{enumerate}

One may ask why we only consider paths to the $100$ most popular websites - what about paths to other IP prefixes?  In this section, we explore such concerns, focusing on questions such as:
\begin{itemize}
\item Do our key ASes also intercept an equally large fraction of paths to other unrelated sites (\emph{e.g.} less popular ones) ?

\item A particular set of users may consider completely different sites ``popular'' (for \emph{e.g.}, users in some countries may only be interested in sites available in their language). Do our key ASes effectively cover paths to such sites?

\item How important are the key ASes to actual censorious nations? If such nations chose to filter paths traversing these key ASes, how would it impact their downstream (foreign) customers?
\end{itemize}

Finally, we discuss the limitations of our method, and our plans for future work.

\subsection{How general are our results?}

Our data shows that a small fraction of ASes ($\approx 30$) cumulatively intercept over $90\%$ of the total paths to popular web destinations (Alexa top 10, 20 ... 200)\footnote{There is a third small caveat: Key ASes cover only more than $90\%$, and not $100\%$ of the paths. But from Houmansadr \cite{houmansadr2014no}, we know that it is not feasible for a country to launch a RAD attack and avoid $90\%$ of paths. The only practical significance is that a user may not get a DR on their first attempt; but if a user probes for a DR, with greater than $99.9\%$ probability she will succeed in three attempts (compared to the $30$ attempts needed for earlier designs \cite{houmansadr2011cirripede}), so this is not a major concern.}. The question naturally follows whether these ASes are specific to the websites chosen for our study, or they intercept a similarly large fraction of all traffic on the Internet.

From Rexford's study of Internet structure \cite{cesareo2012optimizing}, it is reasonable to deduce that there is indeed a small set of ASes - core ASes of the Internet - that cover a majority of routes of the Internet in general. So our task becomes, gathering evidence to show that our $30$ key ASes intercept a large fraction of paths leading to various \emph{other} destinations also.

\begin{figure}[h]
\vspace{-0.75cm}
\includegraphics[scale=0.32]{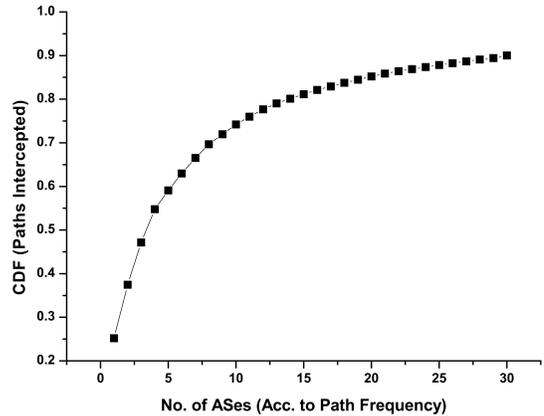}
\vspace{-1.0cm}
\caption{\normalsize{CDF of ASes according to fraction of paths they intercept (for Alexa top-101 to 200 websites).}}
\vspace{-0.5cm}
\label{fig:as-freq-rank_cu_200}
\end{figure}

To begin with, we estimated AS paths to Set B, the set of sites globally ranked $101$--$200$ by Alexa. As Figure~\ref{fig:as-freq-rank_cu_200} shows, the $30$ key ASes identified using paths to Set A, intercepted over $90\%$ of the paths to Set B as well. The same pattern also holds for Set C - sites popular in censorious countries - discussed in the next sub-section.

Finally, we repeated the entire experiment after a gap of four months. We again found the same 30 ASes intercepting over $90\%$ of the paths (see Figure~\ref{fig:as-freq-rank_Censorious_Countries_july2016} in Appendix).

\subsection{How important are the key ASes to actual adversarial nations?}

\begin{figure}[h]
\normalsize
\vspace{-0.8cm}
\includegraphics[scale=0.32]{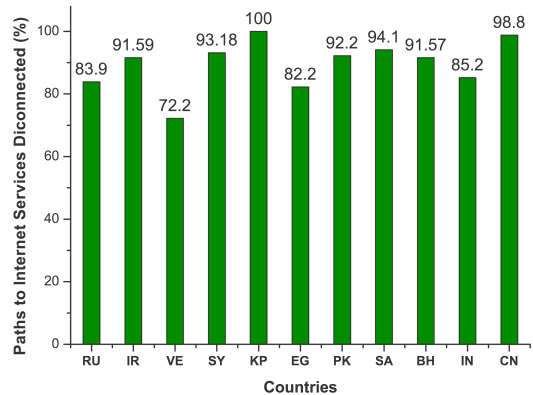}
\vspace{-1.2cm}
\caption{\normalsize{Eleven Censorious Nations: fractions of paths (to major websites) dependent on our $30$ key ASes.}}
\label{fig:ODSP}
\end{figure}

In order to answer this question, we began by measuring how well key ASes cover paths from individual adversary nations to globally important destinations. Our results, showing the fraction of paths disconnected across $11$ censorious nations, are presented in Figure~\ref{fig:ODSP}. The horizontal axis has country names (as 2-letter initials); the vertical axis, the fraction of the paths covered by our key ASes. We see that, for example, our $30$ key ASes cover $98.8\%$ of paths from Chinese ASes to globally popular destinations, and at least $80\%$ for nearly all adversary countries.

However, while these figures are encouraging, they are not enough. For some nations (\emph{e.g.} Iran, China), it might be argued that the loss of paths to globally important sites simply does not matter, as they have their own homegrown substitutes (e.g. \texttt{facenama.com} and \texttt{renren.com} respectively for \texttt{facebook.com}).
 
In response to this concern, we investigated the popular web destinations in censorious countries. As per Alexa \cite{alexaF100}, we find that these include not only local websites, \emph{but also} several of the top-$100$ globally popular sites (search engines, social-media sites, cloud services, e-commerce sites \emph{etc.}). In other words, while the choice of websites does vary across nations (\emph{e.g.} based on user's choice of language), web access is not as ``insular'' as one may fear. 

\begin{figure}[h]
\vspace{-0.75cm}
\centering
\includegraphics[scale=0.32]{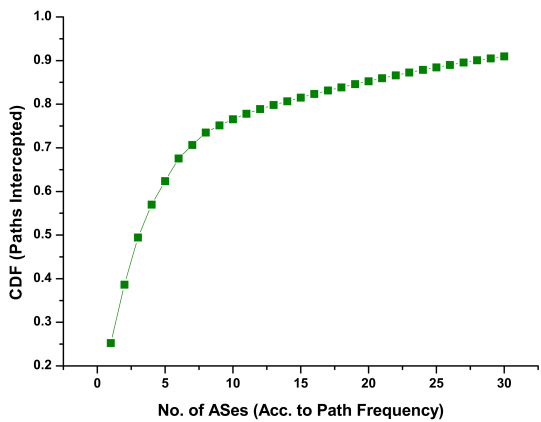}
\vspace{-1.25cm}
\caption{\normalsize{CDF of ASes according to fraction of paths intercepted (for websites popular in censorious nations).}}
\vspace{-0.25cm}
\label{fig:as-freq-rank_Censorious_Countries}
\end{figure}

For each of nine adversary countries studied by Verkamp~\cite{websensor-Foci12} - China, Venezuela, Russia, Syria, Bahrain, Pakistan, Saudi Arabia, Egypt and Iran - we identified our Set C, consisting of the top $50$ websites popular in each of these countries \textbf{(and hosted outside their respective networks)}. Shown in Figure~\ref{fig:as-freq-rank_Censorious_Countries}, our $30$ key ASes intercept $93.3\%$ of the paths originating or transiting these countries and leading to the sites in Set C.

In considering that different paths \emph{originate in} and \emph{transit through} a country, we further realized that avoiding key ASes might be expensive for a country in more ways than one -- \emph{collateral damage}.

\noindent \textbf{Collateral damage:}
Collateral damage results when an AS filters sites, and also causes its customers to lose access \cite{Anonymous:2012:CDI:2317307.2317311}. 
If, for example, China was to boycott the paths routed through our chosen key ASes, Chinese people would lose access to much of the Internet (and certainly to most popular websites); but \emph{so would customers of Chinese ASes}. It becomes a valid question to ask, how many customers are affected?

\begin{figure}[h]
\normalsize
\vspace{-0.75cm}
\centering
\includegraphics[scale=0.32]{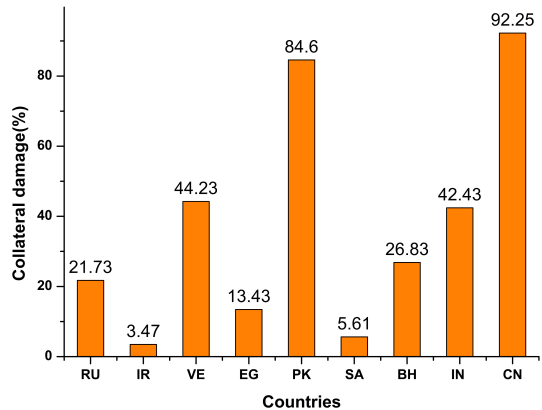}
\vspace{-1.25cm}
\caption{\normalsize{Collateral Damage: Percentage of paths transiting censorious nations that originate at foreign ASes.}}
\vspace{-1.0cm}
\label{fig:collateral_damage}
\end{figure}

To answer this question, we inspected the paths through and from nine censorious countries. Figure \ref{fig:collateral_damage} shows the percentage of paths transiting censorious nations that originate at foreign ASes.

We see that in the case of China, for example, filtering traffic through key ASes would affect a \emph{very large} number of customers, over whom Chinese censorship policies have no control. $306,874$ AS paths, out of a total of $332,742$ paths involving Chinese ASes and leading to popular destinations - \emph{i.e.} $92.25\%$ -  \textit{originate at ASes outside China}\footnote{$362$ particularly interesting paths originated at a Chinese AS, passed through non-Chinese ASes, then re-entered China and passed through Chinese ASes, before finally leaving for their destination.}. 
In fact, our data suggests that collateral damage to customers might be a way to put pressure on several censorious countries; we will explore this in future.

\subsection{Might a different solution do better?}

\textit{The macroscopic analysis (of AS level topology) gives an impression that DR infrastructure is feasible, but the ``devil in the details'' is that the microscopic view (at router level) shows that we would need to convert thousands of routers into DRs.} It is natural to ask whether this conclusion is just an artifact of our method, and whether an alternative approach might find a cheaper solution.

Our approach is not provably optimal. Indeed, we could get by with a smaller number of routers if we placed the DRs to intercept all traffic at a few, fixed overt destinations (\emph{e.g.} Google). However, such a solution is fragile: the censor could simply filter traffic to these overt destinations. Our method of placing DRs uses far fewer ASes than any known comparable methods~\cite{houmansadr2014no}, and intercepts traffic to potential overt destinations (sites that are popular globally and also in censorious nations -- for whom it matters most). Seeing how placing DRs in even our modest number of ASes is a major undertaking, we conclude that there is no ``silver bullet'' -- \emph{robust DR deployment is feasible, but implementing it is a serious challenge}. 

\subsection{Is it easier to cover single countries?}

Our solution involves a single set of ASes that can serve as a DR framework for the overwhelming majority of traffic \emph{globally}. We show that a global DR infrastructure is complex and likely expensive; but might it be feasible to target single censor countries?

We find that in case of major adversaries like China, the best solution is to use the same 30 ASes that we would use for a worldwide DR system. In case of some minor countries such as Syria (which has $2$ ASes), Sorted Ring placement~\cite{houmansadr2014no} does allow a simpler solution: we identified $3$ ASes which intercept all Syrian AS level paths. But the router level maps of these ASes suggest that, even for Syria, we need $1,117$ DRs in 3 different ASes.

Our conclusion is that targeting a DR infrastructure to single countries is difficult even against relatively weak adversaries, and the best solution against strong adversaries (our solution in this paper) is more expensive still.

\subsection{How economically feasible is Decoy Routing?}

Our results show that a comprehensive DR infrastructure would span about $30$ ASes across ten countries, and require massive incentives. The question immediately arises whether existing business models for Decoy Routing~\cite{houmansadr2016game}, \emph{i.e.} \emph{central deployment} (where a single organization pays individual AS operators to deploy DRs) and \emph{autonomous deployment} (where ASes individually deploy DRs and bill their users for usage), can reasonably provide such incentives.

In the case of central deployment, we note that unlike, for instance, Tor, this project will depend on large-scale corporate participation. Tor is a globally distributed \emph{volunteer} network, involving participants running the open-source Tor software on their (personal) end-hosts; the actual funding for the project only needs to support the developers, maintainers and some minimal infrastructure (\emph{Directory Authority} servers, etc.) A worldwide DR framework needs to incentivize multiple multi-billion dollar companies to co-operate, and it is disturbingly likely that a single player who pays such incentives - whether a major company or a government - is motivated by its own agenda, rather than benevolence.

Autonomous deployment suffers from an even more serious issue. Decoy Routing obfuscates public knowledge of the deployment infrastructure (physical location and hosting network); \emph{but such obfuscation also makes it difficult for users to target payment}. Any Internet based payment scheme would reveal the identities of DR hosting ASes to the clients, and in time, to their censorious ISPs. The adversary now simply blocks such Internet based payment transactions in order to prevent users from getting DR service; the whole ``robust infrastructure that cannot be routed around'' is rendered moot.

We therefore conclude that a practical DR infrastructure faces substantial challenges, and would likely only be possible with major support from one or more powerful nations. 

\subsection{Methods, Limitations, and Future Work.}

This sub-section is devoted to the choices we made, w.r.t. the design of our methods of network mapping. We explain our choices, their limitations, and how we propose to go forward in future.

\noindent \textbf{Choice of AS:}
The first major question in our study, was how to choose key ASes. It may be argued that we could simply have chosen Tier-1 ASes -- \emph{i.e.} ASes that have no provider -- or ASes with the largest customer cone size (The customers, customers of customers, \emph{etc.} of an AS are said to form its ``customer cone''.), based on publicly available data~\cite{caida-inter-as-rels}, as proposed earlier~\cite{houmansadr2014no}.
Why pick key ASes by path frequency? It turns out that there is indeed a good reason for directly choosing ASes by path frequency, \emph{i.e.} by how many of the paths they intercept. 

\begin{figure}[h]
\centering
\normalsize
\vspace{-0.25cm}
\includegraphics[width=8cm]{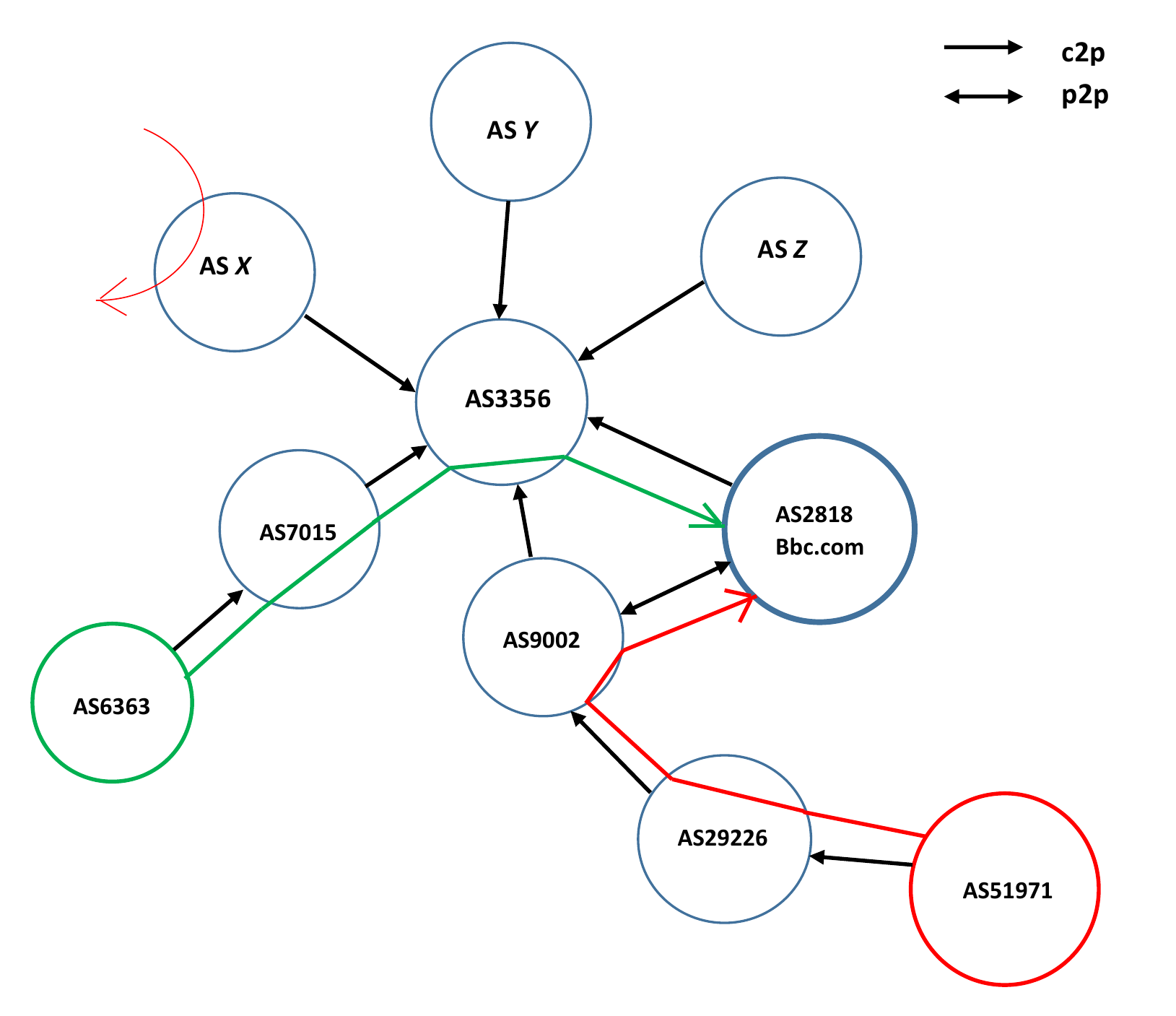}
\vspace{-0.25cm}
\caption{\normalsize{Valley free paths in the cone of AS3356. Green line: network path traversing AS3356 to reach AS2818 directly. Red lines: network path through one-hop customers of AS3356, but not AS3356 itself.}}
\label{fig:BBC}
\vspace{-0.25cm}
\end{figure}

A substantial fraction of AS paths traverse the customers of \emph{``root''} ASes(\emph{i.e.}, those with very large customer cones) without traversing the root ASes themselves. For \emph{e.g.}, the traffic through AS9002 to AS2818 (\texttt{www.bbc.co.uk}) does not pass through AS3356, though it is the provider to both these ASes (see Figure~\ref{fig:BBC}).

Unsurprisingly, $34.16\%$ of the paths to top-$100$ IP-prefixes traverse the AS with the largest customer cone, AS3356 (cone size = $24,553$). But nearly as many paths, $33.17\%$, pass through its 1-hop (immediate) customers, and not AS3356. In the Appendix, we present more such figures (Table~\ref{tab:first-hop-unreachable}), and show that Spearman's Rank Correlation coefficient \cite{gautheir2001detecting} between AS ranks by path frequency and by cone size is only $0.2$. Given the considerable fraction of paths which do not transit root ASes with large cone sizes (preferring to transit through their customer ASes instead), we conclude that \emph{customer-cone size is not a good parameter to choose key ASes}.

\noindent \textbf{AS path estimation:}
The two main methods of estimating an inter-AS topology are: (a) using \texttt{traceroute} traces (as in CAIDA Ark) (b) using BGP routing tables. Traceroute data, being constrained by the location of available probing nodes, is not sufficiently rich to estimate the actual path of traffic from every AS to a given prefix. Hence we choose the routing-table approach. 

Previous efforts use simulated BGP paths ~\cite{houmansadr2014no}, or paths derived from a Breadth-First traversal of inter-AS links \cite{cesareo2012optimizing}. We improve upon this by employing Gao's algorithm with \emph{real} BGP tables (collected from various Internet Exchanges~\cite{routeviews}), thereby estimating the \emph{actual} paths from a chosen IP prefix to all ASes (at a given point of time).


Of course, our map is still not perfect. As Gregori \emph{et al.} ~\cite{gregori2016isolario,gregori2015novel} show, publicly-available routing tables have biases, errors and bogus route advertisements. In order to address such issues, in future, we may cross-validate our map with different sources of data (and mapping algorithms).

\noindent \textbf{Router level topology estimation:}
Router-level mapping of AS structure, uses \texttt{traceroute} probes from various \texttt{planetlab} hosts to IP addresses inside the ASes.  We are limited by the fraction of routers discoverable through \texttt{traceroute} probes, and routers and middleboxes are sometimes set up to not respond to \texttt{ping} and \texttt{traceroute}. To limit this concern as far as possible, we use \texttt{Paris Traceroute}, with TCP probes. Secondly, we used \texttt{planetlab} nodes to launch \texttt{traceroute} probes, as the \texttt{looking-glass} servers (as used originally~\cite{rocketfuel-paper}) were unavailable at the time of our tests. There is always a chance some routes are simply not covered; increasing the number of probing hosts may improve our topology estimation, as new paths may be discovered by probing an IP address from different vantage points. 



\section{Concluding Remarks}
\label{sec:conclusion}

In this paper, we have made two contributions towards answering the question of how to best place DRs in the Internet. 

\begin{enumerate}
\item As our first contribution, we demonstrate that a small set of candidate ASes ($\approx 30$) intercepts a very large fraction of paths (greater than $90\%$) to sites of interest, \emph{i.e.} potential overt destinations, \emph{irrespective of the adversary country}. In other words, placing DRs in our ASes is sufficient to build a global DR framework. (As opposed to current approaches \cite{houmansadr2014no, houmansadr2016game}, which need the collaboration of over $800$ ASes for a single adversary such as Venezuela or China.). We also observe that, if censorious regimes (like China) attempt to filter traffic along the paths transiting our $30$ ASes, they will not only censor their own citizens, but many other residing outside their network boundaries (collateral damage).



\item Our second contribution is to explore the question of DR placement, not only at the AS-level, but at the router level. In practice, an AS is not a simple entity; it may have thousands of routers, and it is not obvious which of these should be replaced with DR. We find that, to intercept a large fraction of paths through an AS, we need a large number of both edge and core routers - typically several hundred (and in cases such as Quest Communications and Verizon, well over $1500$ routers).
\end{enumerate}

Thus, setting up a worldwide DR framework may require the collaboration of a small set of ASes ($\approx 30$). But even a single key AS, on an average, will need several hundred routers to intercept all the paths. We conclude that building a worldwide DR infrastructure is practically feasible, \emph{but} ASes need sufficiently strong incentives to deploy a total of over $11,700$ DRs. We will explore such issues, in our future work.

\section*{Acknowledgements}
We would like to express thanks to Mr. Rahul Singh,  M.Tech IIITD, who helped us develop our first  prototype of Gao~\emph{et al.}'s AS mapping algorithm. We are also grateful to Persistent Systems Ltd., India, who very kindly agreed to fund our travel to the conference to present the paper.

\begin{small}
\bibliographystyle{IEEEtran}
\bibliography{2046730.bib}
\appendix

\section{Gao AS-level mapping:} 
\label{sec:app_gao}

Our end-to-end AS-level path inference algorithm follows Gao \cite{qiu2006cam04}. The key idea of this approach is to construct paths based on the routing information in existing BGP routing tables. More precisely, the process estimates paths from an IP-prefix to every other AS of the Internet.


 
The inputs to the algorithm are existing BGP RIBs, collected by the RouteViews project~\cite{routeviews} from Internet Exchange Points (IXes), where several ASes peer.

Paths directly obtained from RIBs are termed \emph{sure paths}. ASes on sure paths are called \emph{Base ASes}. For example, in the (hypothetical) path $2869-3586-49561-58556-10348$ $192.168.1.12/24$, each number represents an AS. The path originates at AS2869 and terminates at AS10348, the home AS of the advertised prefix $192.168.1.12/24$. Note that the suffixes of sure paths are themselves also sure paths. 

In addition to sure paths, the algorithm computes new ones. This is done by extending sure paths to other ASes to which there are no explicitly-known paths (from the prefix concerned). The extended path must be loop-free, and must satisfy the \emph{Valley-Free Property}~\cite{gao2001inferring}. The process is as follows.
\begin{itemize}
\item For each prefix, all sure paths (containing all the base ASes) are selected. (These are simply the RIB entries corresponding to the input prefix). 

Next, these sure paths are to be inspected for possible extension to new ASes, provided they they satisfy the Valley Free property and have no loops.

\item The algorithm searches for ASes in existing RIBs that share valid business relations with the end ASes of paths. [We used the relationships presented by CAIDA ~\cite{caida-inter-as-rels}.]

\item An edge is \emph{assumed}, to extend a sure path by one hop. 

Note that we are trying to find a path from an AS to the target prefix, and that extensions of several sure paths might connect the chosen AS to the prefix. Hence there is a need for tie breaking.
\begin{itemize}
\item  The algorithm sorts the possible paths, and selects the \emph{shortest} path to the prefix. 

\item In case of a tie, the path with minimum \emph{uncertainty} (length of the inferred path extensions) is chosen. 

\item If there is still a tie, the path with the higher \emph{frequency index} (the number of times a sure path actually appears in the RIBs) is selected.
\end{itemize}

\item Appropriate data structures, \emph{i.e.} the frequency with which an edge appears in the RIBs, the uncertainty of the extended path, and the new path length, are updated.
\end{itemize}

\section{Path frequency vs customer-cone size}

\begin{figure}[h!]
\centering
\normalsize
\includegraphics[scale=0.5]{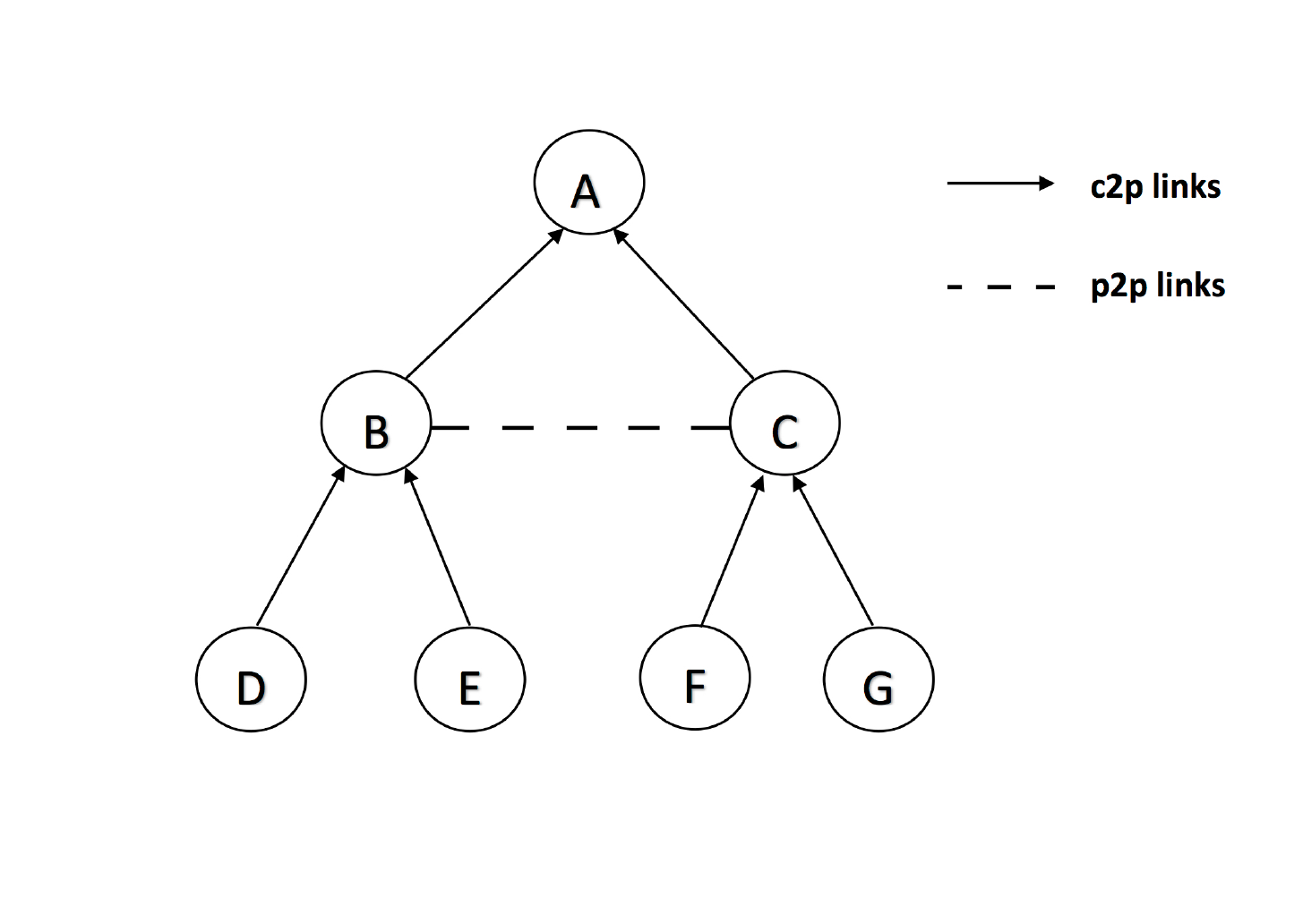}
\vspace{-0.5cm}
\caption{\normalsize{Schematic AS graph with multiple valid valley-free paths: $D-B-E$, $D-B-C-F$, $D-B-C-G$, $D-B-A-C-F$, $D-B-A-C-G$, $E-B-A-C-F$ and $E-B-A-C-G$. Note how some do not traverse $A$, the AS with the highest customer-cone size.}}
\label{fig:valley_free_valid}
\vspace{-0.5cm}
\end{figure}

We provide some detail for our claim in Section 6.4, that customer-cone size is not a reliable metric to identify the ASes that transport a large fraction of traffic. We explain our reasoning with the example of the AS graph in Figure ~\ref{fig:valley_free_valid}.

The figure represents a hypothetical AS graph where node $A$ represents an AS with the highest customer-cone size of $6$, the total number of ASes that A can reach via its customers and their customers ($D,B,E,F,C,G$). ASes $B$ and $C$ have customer cones of size $2$ (for each of the individual nodes).

\begin{table}[b!]
\centering
\normalsize
 \begin{tabular}{|l|l|l|} 
 \hline
\textbf{ASN} & \textbf{\% of path} & \textbf{\% of path}\\
             & \textbf{not reaching} & \textbf{reaching}\\
             & \textbf {the AS}& \textbf {the AS}\\\hline  
3356 & 34.16 & 33.17\\\hline
174 & 29.05 & 13.13 \\\hline
2914 & 28.16 & 12.90  \\\hline
1299 & 36.50 & 8.05 \\\hline
3257 & 21.00 & 5.23 \\\hline
6939 & 7.46 & 4.40\\\hline
6461 & 5.13 & 4.03\\\hline
6453 & 26.00 & 3.76 \\\hline
7018 & 7.40 & 3.70\\\hline
10310 & 0.07 & 3.52\\\hline
\end{tabular}
\vspace{3mm}
\caption{\normalsize{Prefix-to-AS paths in cone of core ASes: \%age traversing core ASes themselves, vs. \%age traversing their immediate ($1$-hop) customers.}}
\vspace{-0.5cm}
\label{tab:first-hop-unreachable}
\end{table}

There are several valid valley free paths in this hypothetical AS graph: $D-B-E$, $D-B-C-F$, $D-B-C-G$, $D-B-A-C-F$, $D-B-A-C-G$, $E-B-A-C-F$ and $E-B-A-C-G$. However, as evident from the example, not all of them pass through the \emph{root AS}, \emph{i.e.} the node with the highest customer-cone size. This is also evident from Table ~\ref{tab:first-hop-unreachable}: for several large ASes, a considerable fraction of paths do not traverse the core ASes themselves, but do traverse their immediate ($1$-hop) customers \footnote{Interestingly, smaller customer-cones have fewer such paths, i.e. paths not traversing the root AS. Perhaps smaller cones have fewer neighbor ASes to route through?  We may study this in future work}.

In fact, customer cone size is not even well correlated with path frequency. The Spearman's Rank Correlation Coefficient is only $\approx 0.2$ (see Figure \ref{fig:Spearman}).

\begin{figure}[h!]
\centering
\normalsize
\vspace{-0.5cm}
\includegraphics[scale=0.32]{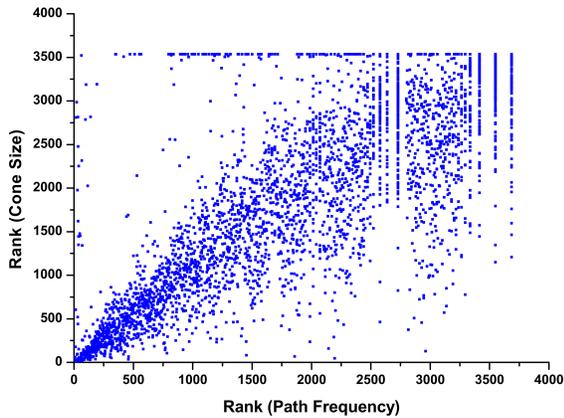}
\vspace{-1.0cm}
\caption{\normalsize{AS Rank variation: path frequency vs cone size for transit ASes.}}
\label{fig:Spearman}
\end{figure}

\section{Additional Graphs}

\noindent \textbf{Traffic to specific destinations:}

\begin{figure}[h!]
\vspace{-0.5cm}
\includegraphics[scale=0.32]{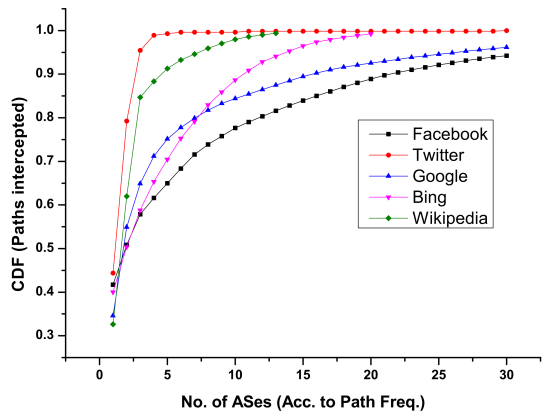}
\vspace{-0.8cm}
\caption{\normalsize{
CDF of ASes according to fraction of paths to popular websites that they intercept}}
\label{fig:as-freq-rank_cu_fb}
\end{figure} 

For completeness, we present the results of our experiment for a few of the most important single websites in isolation, in Figure \ref{fig:as-freq-rank_cu_fb}.

Clearly, while single websites are far more variable, the general trend is similar to Figure~\ref{fig:asfreqrankcu100}. About $15$ ASes (out of the $50$ heavy-hitter ASes identified) cover over $80\%$ of the AS-paths to these destinations. 

Only $5$ ASes collectively transport all the paths to the prefix corresponding to \texttt{twitter.com}, while about $18$ ASes intercept all paths carrying traffic from \texttt{bing.com}. Finally, about $30$ ASes cumulatively transport about $98\%$ of the paths corresponding to \texttt{google.com} and \texttt{facebook.com}.

For comparison, here is the cumulative frequency of the paths intercepted by ASes, this time for paths to the Alexa top-200 sites.

\begin{figure}[h!]
\includegraphics[scale=0.32]{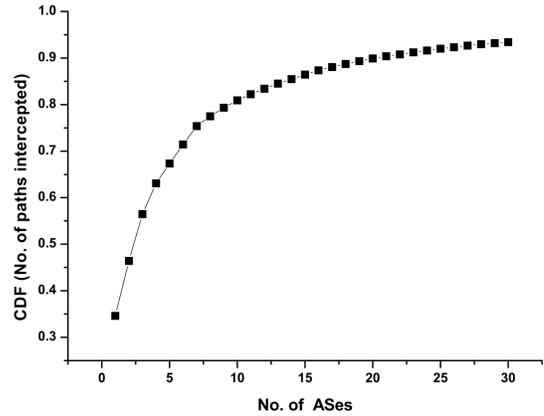}
\vspace{-1.25cm}
\caption{\normalsize{CDF of ASes by fraction of paths that they intercept (for Alexa top-200 sites)}}
\label{fig:as-freq-rank_Censorious_Countries_july2016}
\end{figure}

We also provide the graph of the cumulative path coverage inside an AS, by its heavy hitter routers. As is quite clear, the graph varies a good deal; some ASes are almost completely covered by a few routers, while others (AS 3257) need very many heavy-hitters, and can more easily be covered by choosing their edge routers (Figure \ref{fig:Rocket_top_five}).

\begin{figure}[h!]
\centering
\vspace{-0.75cm}
\includegraphics[scale=0.32]{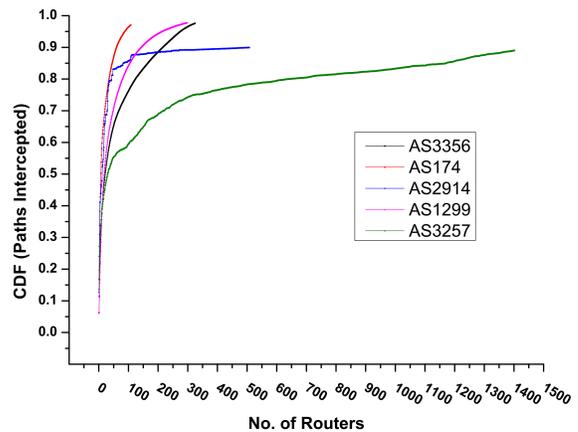}
\vspace{-1.25cm}
\caption{\normalsize{\texttt{Traceroute} paths in the top five (out of 30) key ASes. The number of routers needed to cover $90\%$ of the paths varies between $288$ (AS174) and $1483$ (AS3257)}}
\label{fig:Rocket_top_five}
\end{figure}

\end{small}

\end{document}